\documentclass[runningheads]{llncs}

\ifdefined\pdfpagewidth
\fi
\usepackage[T1]{fontenc}
\usepackage{graphicx}
\usepackage{amsmath}
\usepackage{amssymb}
\usepackage{mathtools}
\usepackage{multirow}
\usepackage{booktabs}
\usepackage{csquotes}
\usepackage[table]{xcolor}
\usepackage{adjustbox}
\usepackage{pifont}
\usepackage{url}
\usepackage{bbding}
\usepackage{hyperref}
\hypersetup{hidelinks}

\newcommand{\figvspace}{\vspace{-0.3em}}
\newcommand{\tabvspace}{\vspace{-0.2em}}

\definecolor{headergray}{HTML}{E8E8E8}     
\definecolor{lightblue}{HTML}{E8E8E8}      
\definecolor{tableheadercolor}{HTML}{E8E8E8}
\definecolor{lightyellow}{HTML}{FFF7E6}    
\definecolor{ourscolor}{HTML}{FFF7E6}      
\definecolor{lightgreen}{HTML}{D4EDDA}
\definecolor{lightred}{HTML}{F8D7DA}
\definecolor{dropred}{rgb}{0.8, 0, 0}
\definecolor{improvered}{rgb}{0.8, 0, 0}
\definecolor{bestc}{HTML}{F0F0F0}          
\definecolor{secondc}{HTML}{F0F0F0}        

\newcommand{\best}[1]{\textbf{#1}}
\newcommand{\second}[1]{\underline{#1}}

\begin{document}

\title{Anatomy of a Decision: Uncertainty-aware Hierarchical Intent Learning via Flow Matching for Multimodal Recommendation}
\titlerunning{Uncertainty-aware Hierarchical Intent Learning}

\author{Yuchen Miao\inst{1} \and Zijun Wang\inst{1} \and Ke Liu\inst{1} \and Siyang Xu\inst{2,3}\textsuperscript{\Envelope}}
\authorrunning{Y. Miao et al.}

\institute{Sydney Smart Technology College, Northeastern University, China\\
\and
School of Computer and Communication Engineering and Hebei Key Laboratory of Marine Perception Network and Data Processing, Northeastern University at Qinhuangdao, China
\and
Shijiazhuang Innovation Research Institute of Northeastern University, Shijiazhuang, China\\
\textsuperscript{\Envelope}Corresponding author.}

\maketitle
\thispagestyle{headings}

\begin{abstract}
Modeling the underlying user intent is crucial for recommendation, but existing methods struggle with the inherent uncertainty and the dynamic, hierarchical nature of user interests. Current approaches often rely on clustering or prototype learning to discover a static set of intents. However, they face two critical challenges: (1) they overlook the uncertainty inherent in multimodal features; and (2) they assume a static and flat intent structure, failing to adapt to a user's varying decision certainty. To address these limitations, we propose \textbf{UHIFlow}, an \textbf{U}ncertainty-aware \textbf{H}ierarchical \textbf{I}ntent learning framework via \textbf{Flow} matching. First, our Cross-modal Uncertainty Synergistic Modeling (CUSM) module leverages conditional flow matching to quantify uncertainty from visual and textual modalities and synergistically align them. Subsequently, the Uncertainty-guided Hierarchical Intent Generation (UHIG) module uses this quantified uncertainty to dynamically construct a personalized intent hierarchy, generating coarse-grained intents for uncertain users and fine-grained ones for users with clear preferences. Extensive experiments on three real-world datasets demonstrate that UHIFlow significantly outperforms state-of-the-art baselines.

\end{abstract}

\section{Introduction}
Recommender systems are indispensable for navigating online information across domains from e-commerce to content streaming. They model user preferences from past behaviors, yet raw behaviors are only superficial signals of deeper, underlying intents. To bridge this gap, intent-based recommenders have emerged as a promising frontier~\cite{jannach2024survey,liu2024end,miao2026r3rec}.

Existing intent-based recommendation falls into two categories. The first clusters user behaviors to discover intents, grouping users by interaction patterns. The second employs prototype learning, mapping users to a latent space of predefined intent prototypes that represent canonical interests~\cite{liu2024end,ren2023disentangled,zhang2024exploring}. More recently, learning intent from items' visual and textual information has gained traction, offering a more holistic view of preferences~\cite{yang2024multimodal}.

\begin{figure}[h]
     \centering
     \includegraphics[width=0.7\linewidth]{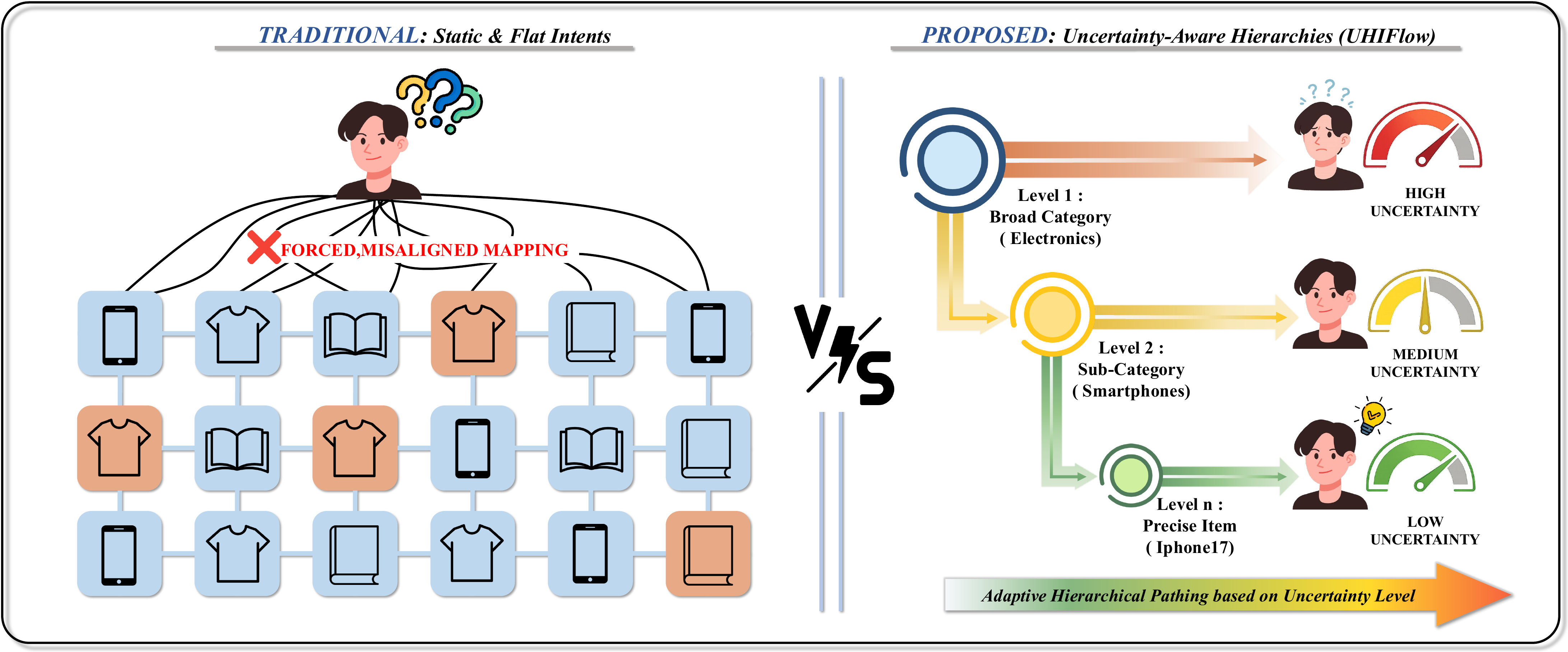}
     \caption{A comparison between traditional intent modeling and our proposed UHIFlow. The traditional approach (left) relies on static, flat intents, often leading to \textbf{forced, misaligned mappings} across items of \textbf{distinct categories (indicated by different colors)}. In contrast, UHIFlow (right) employs \textbf{adaptive hierarchical pathing} guided by uncertainty levels. It maps users with \textbf{High Uncertainty} to broad categories (Level 1, e.g., ``Electronics'') and progressively refines the representation for users with \textbf{Low Uncertainty} down to precise items (Level $n$, e.g., ``iPhone 17'').}
     \label{fig:motivation}
\figvspace
\end{figure}

Despite these advancements, existing multimodal intent modeling approaches face the following critical challenges that limit their effectiveness in real-world scenarios as shown in figure~\ref{fig:motivation}:
\begin{itemize}
    \item \textbf{Neglect of Inherent Uncertainty.} They overlook the uncertainty in recommendation, which stems both from ambiguous user behavior and from the multimodal data itself, e.g., noisy images or semantically ambiguous text~\cite{gao2024embracing}. Treating multimodal features as deterministic inputs prevents a reliable representation of user intent, yielding suboptimal performance.
    \item \textbf{Static and Flat Intent Structure.} They assume a fixed number of intents, ignoring the hierarchical nature of user intent. Decision-making is often a coarse-to-fine refinement: a user with high uncertainty (\enquote{just browsing}) holds a coarse intent (\enquote{electronics}), whereas one with low uncertainty (\enquote{looking for a new phone}) holds a much finer intent. A static structure cannot adapt to this varying certainty.
\end{itemize}

To address these challenges, we propose \textbf{U}ncertainty-aware \textbf{H}ierarchical \textbf{I}ntent learning via \textbf{Flow} matching (\textbf{UHIFlow}). Since flow matching learns complex distributions without restrictive priors~\cite{lipman2022flow}, we treat item features as distributions rather than fixed points to model their uncertainty explicitly. We first introduce a Cross-modal Uncertainty Synergistic Modeling module that quantifies the uncertainty of visual and textual modalities and lets the two uncertainties inform and regularize each other. We then design an Uncertainty-guided Hierarchical Intent Generation module that uses this uncertainty to construct a user-specific intent hierarchy, generating coarse-grained structures for uncertain users and fine-grained ones for users with clear preferences.

The main contributions of this paper are summarized as follows:
\begin{itemize}
    \item We propose UHIFlow for uncertainty-aware multimodal intent modeling; to our knowledge, it is the first to explicitly model multimodal uncertainty for hierarchical intent discovery in recommendation.
    \item We design a Cross-modal Uncertainty Synergistic Modeling module that leverages conditional flow matching to quantify and synergize uncertainties from different modalities.
    \item We devise an Uncertainty-guided Hierarchical Intent Generation module that adaptively constructs personalized intent hierarchies based on user uncertainty.
    \item Extensive experiments on real-world datasets show that UHIFlow outperforms state-of-the-art baselines.
\end{itemize}

\section{Related Work}

\subsection{Multimodal Recommendation}
Multimodal recommendation incorporates visual and textual information to enrich collaborative signals. Early work such as VBPR~\cite{he2016vbpr} demonstrates the benefit of visual features, while recent methods exploit stronger cross-modal learning: MMSSL~\cite{wei2022mmssl} aligns multimodal and collaborative signals, DiffCL~\cite{song2025diffcl} constructs contrastive views, LGMRec~\cite{guo2024lgmrec} unifies local and global graph learning, and DiffMM~\cite{jiang2024diffmm} introduces graph diffusion. MMIL models multimodal intentions with a learnable codebook, while MDN and Diff-MSIN decompose or synergize modality factors~\cite{yang2024multimodal,liu2025mdn,cui2025diffusion,shi2026motif}. However, these methods rarely quantify modality uncertainty or adapt the hierarchy of user intent.

\subsection{Intent Modeling in Recommender Systems}
Intent modeling captures multi-faceted user preferences that single embeddings often miss~\cite{jannach2024survey}. Existing methods include contrastive or clustering-based intent discovery~\cite{chen2022intent,liu2024end}, multi-interest modeling with MIND~\cite{li2019mind}, and disentangled factor learning~\cite{ren2023disentangled,zhang2024exploring}. Hierarchical formulations such as HIM~\cite{Zhu2022HIM}, HieRec~\cite{qi2021hierec}, and recent hierarchical or semantic-token generative recommendation~\cite{chen2026hierarchical,hu2026ids} encode coarse-to-fine preferences. Yet most rely on fixed intent numbers or static structures, limiting their ability to adjust intent granularity according to user certainty.

\subsection{Generative Models for Recommendation}
Generative recommendation has advanced through diffusion and flow-based models. DiffRec~\cite{WangXFL0C23} and Diff4Rec~\cite{Wu0CLHS023} apply diffusion to interaction modeling and sequential augmentation, while recent diffusion models denoise multimodal features and behaviors~\cite{song2025diffcl,jiang2024diffmm,lu2025dmmd4sr,cui2025multi}. Flow matching offers an efficient way to learn complex distributions without restrictive priors~\cite{lipman2022flow}, and FlowCF~\cite{liu2025flow} adapts it to collaborative filtering. These works focus mainly on denoising or generation rather than using flows to quantify multimodal uncertainty for hierarchical intent learning.

\subsection{Uncertainty Modeling in Recommendation}
Uncertainty modeling improves robustness under sparsity, distribution shift, and cold-start scenarios. UCC~\cite{Liu2023UCC} uses teacher-student consistency, UA\mbox{-}GNN~\cite{Cao2024UAGNN} performs uncertainty-aware pseudo-label selection, and WaPOIR~\cite{Zhou2023WaPOIR} learns Wasserstein embeddings. Multimodal fusion research also shows that modalities provide unequal and unreliable evidence~\cite{gao2024embracing}. Existing methods, however, seldom combine multimodal uncertainty estimation with adaptive hierarchical intent generation.

\section{Preliminary}

In a typical multimodal recommendation scenario, we have a set of users $\mathcal{U}$ and a set of items $\mathcal{I}$. The user-item interactions are represented by a matrix $\mathbf{R} \in \mathbb{R}^{|\mathcal{U}| \times |\mathcal{I}|}$, where $r_{ui}=1$ if user $u$ has interacted with item $i$, and $r_{ui}=0$ otherwise. Each item $i \in \mathcal{I}$ is associated with multimodal features, specifically a visual feature vector $\mathbf{v}_i \in \mathbb{R}^{d_v}$ and a textual feature vector $\mathbf{t}_i \in \mathbb{R}^{d_t}$.

The objective is to learn a user representation $\mathbf{u}_u^*$ and an item representation $\mathbf{h}_i$ for predicting a personalized ranked list, with the recommendation score $\hat{y}_{ui} = (\mathbf{u}_u^*)^\top \mathbf{h}_i$.

\section{Methodology}

We first describe the multimodal interest modeling layer, then the two core modules---Cross-modal Uncertainty Synergistic Modeling (CUSM) and Uncertainty-guided Hierarchical Intent Generation (UHIG)---and finally the uncertainty-aware aggregation and overall optimization objective.


\subsection{Multimodal Interest Modeling Layer}

\subsubsection{User--Item Interaction Graph}
We construct a bipartite graph $\mathcal{G} = (\mathcal{V}, \mathcal{E})$ where $\mathcal{V} = \mathcal{U} \cup \mathcal{I}$ is the set of user and item nodes, and $\mathcal{E}$ represents the observed interactions. The initial feature for an item node $i$ is derived by concatenating and projecting its visual and textual features: $\mathbf{e}_i^{(0)} = \text{MLP}([\mathbf{v}_i; \mathbf{t}_i])$. User nodes are initialized with learnable embeddings $\mathbf{e}_u^{(0)}$.

\subsubsection{Collaborative Interest Propagation}
To inject high-order collaborative information, we employ a light-weight graph propagation mechanism similar to LightGCN \cite{he2020lightgcn}. The embeddings are refined over $K$ layers as follows:
\begin{equation}
\mathbf{e}_u^{(k)} = \sum_{i \in \mathcal{N}_u} \frac{1}{\sqrt{|\mathcal{N}_u||\mathcal{N}_i|}} \mathbf{e}_i^{(k-1)}, \quad \mathbf{e}_i^{(k)} = \sum_{u \in \mathcal{N}_i} \frac{1}{\sqrt{|\mathcal{N}_i||\mathcal{N}_u|}} \mathbf{e}_u^{(k-1)}
\end{equation}
The final user and item embeddings, $\mathbf{h}_u$ and $\mathbf{h}_i$, are obtained by aggregating the embeddings from all layers: $\mathbf{h}_u = \sum_{k=0}^K \alpha_k \mathbf{e}_u^{(k)}$ and $\mathbf{h}_i = \sum_{k=0}^K \alpha_k \mathbf{e}_i^{(k)}$, where $\alpha_k$ are layer-wise aggregation weights. These embeddings serve as the basis for the subsequent uncertainty and intent modeling.

\begin{figure}[h]
    \centering
    \includegraphics[width=0.95\textwidth]{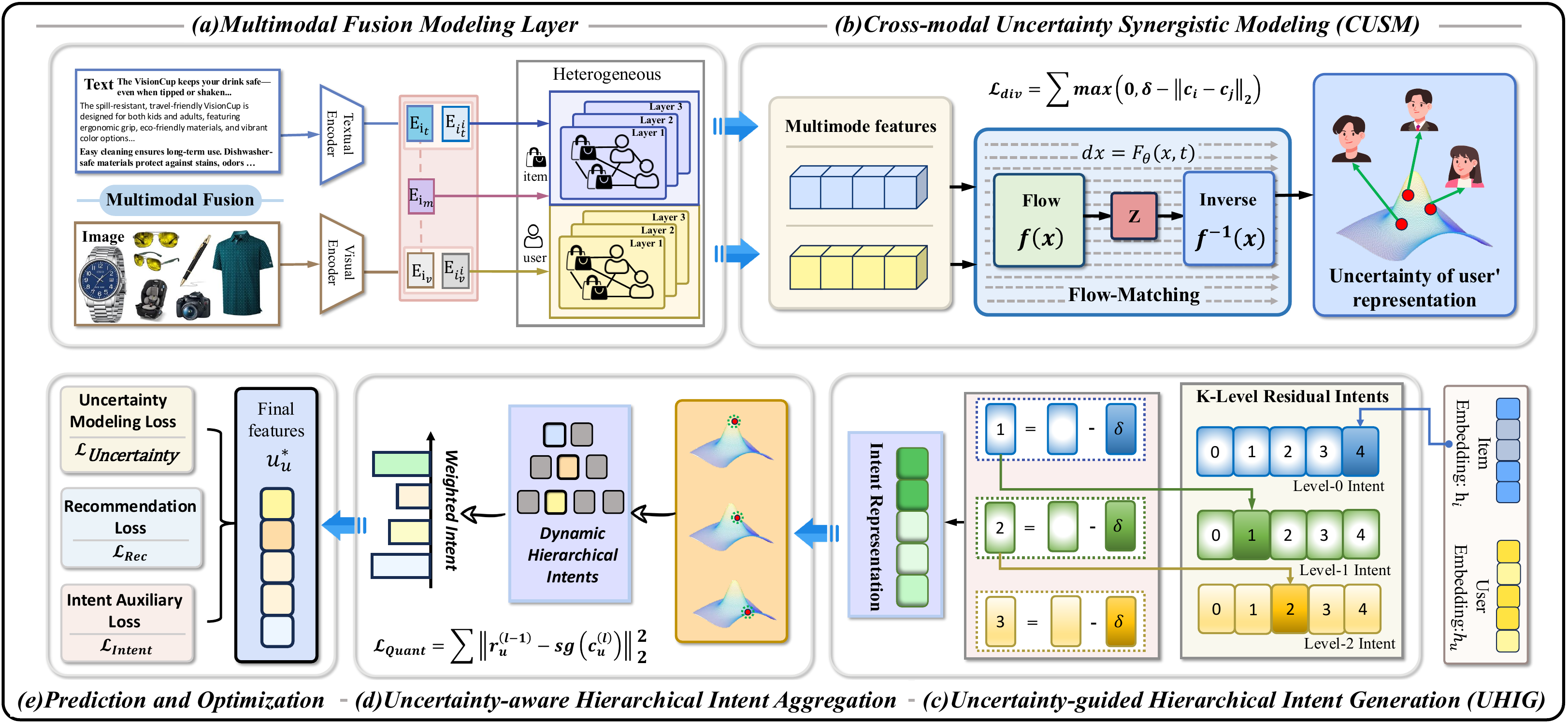}
    \caption{Overview of UHIFlow, including multimodal interest modeling, CUSM, UHIG, and uncertainty-aware aggregation for final recommendation.}
    \label{fig:model}
\figvspace
\end{figure}

\subsection{Cross-modal Uncertainty Synergistic Modeling (CUSM)}
This module models item features as rich probability distributions rather than deterministic point estimates. Since the true conditional distribution of an item's multimodal features, $p_{data}(\mathbf{x}|\mathbf{c})$, is complex and uncertain, we learn a model $q_{\theta}(\mathbf{x}|\mathbf{c})$ that captures this uncertainty.

We select Flow Matching \cite{lipman2022flow} for this task due to its ability to learn complex, data-dependent distributions without imposing restrictive prior assumptions. To make modality-specific uncertainty modeling explicit, we instantiate two conditional continuous normalizing flows (CNFs), $f_\theta$ and $f_\phi$, for visual and textual features. The visual flow $f_\theta$ is induced by the vector field $v_\theta$, while the textual flow $f_\phi$ is induced by $v_\phi$. For modality $m \in \{v,t\}$, the corresponding flow evolves according to
\[
\frac{d\mathbf{x}_m(t)}{dt}=v_m(t,\mathbf{x}_m(t),\mathbf{h}_i),
\]
where $v_v \equiv v_\theta$ and $v_t \equiv v_\phi$. Each flow transforms a simple base distribution $p_0(\mathbf{x}) = \mathcal{N}(\mathbf{0}, \mathbf{I})$ into the modality-specific target feature distribution conditioned on the item's collaborative embedding $\mathbf{h}_i$.

Let $\mathbf{z}_v, \mathbf{z}_t$ be latent variables for visual and textual features. The flow matching objective trains each vector field $v_m(t, \mathbf{x}, \mathbf{c})$ to match a target field defined by a probability path $p_t(\mathbf{x}|\mathbf{x}_1)$ connecting the noise distribution $p_0$ to the data distribution $p_1$. Throughout this section, the condition is $\mathbf{c} = \mathbf{h}_i$, while the target sample is drawn from $p(\mathbf{x}|\mathbf{v}_i)$ (or $p(\mathbf{x}|\mathbf{t}_i)$). The conditional flow matching losses for the visual and textual modalities are aggregated as:
\begin{equation}
\mathcal{L}_{FM} = \mathbb{E}_{p(\mathbf{h}_i, \mathbf{v}_i, \mathbf{t}_i)} \left[ \mathcal{L}_{FM_v}(\theta) + \mathcal{L}_{FM_t}(\phi) \right]
\end{equation}
where, specifically, the visual loss is defined as:
\begin{equation}
\mathcal{L}_{FM_v}(\theta) = \mathbb{E}_{t, p_t(\mathbf{x}|\mathbf{v}_i)} \left[ \| v_\theta(t, \mathbf{x}, \mathbf{h}_i) - u_t(\mathbf{x}|\mathbf{v}_i) \|^2 \right]
\end{equation}

This objective has a distributional interpretation: reducing the gap between the learned vector field and the target probability path transports the base distribution toward the data-conditioned feature distribution. The resulting density evolution provides both denoised multimodal representations and a principled signal for uncertainty estimation.

To keep uncertainty estimation coherent across modalities, we introduce a synergistic guidance mechanism: since ambiguity in one modality should inform the other, we align the intermediate latent representations of the two flows. Let $\Phi_s(\mathbf{z}; f, \mathbf{h}_i)$ denote the state at time $s$ obtained by evolving noise $\mathbf{z}$ with flow $f$ conditioned on $\mathbf{h}_i$. The cross-modal guidance loss is:
\begin{equation}
\mathcal{L}_{cross} = \mathbb{E}_{s \sim U(0,1), \mathbf{z}_v, \mathbf{z}_t \sim \mathcal{N}} \| \Phi_s(\mathbf{z}_v; f_\theta, \mathbf{h}_i) - \text{Proj}(\Phi_s(\mathbf{z}_t; f_\phi, \mathbf{h}_i)) \|_2^2
\end{equation}
where $\text{Proj}(\cdot)$ is a linear projection for modality alignment. Finally, we quantify the uncertainty of each modality via the magnitude of the log-density change, computed using the instantaneous change of variables formula:
\begin{equation}
\varsigma_m(\mathbf{h}_i) = \mathbb{E}_{\mathbf{z}_m \sim \mathcal{N}} \left[ \left| \int_0^1 \text{Tr}\!\left(\nabla_{\Phi_s} v_m(s, \Phi_s(\mathbf{z}_m; f_m, \mathbf{h}_i), \mathbf{h}_i)\right) ds \right| \right]
\end{equation}
where $m \in \{v, t\}$. The absolute value ensures $\varsigma_m \ge 0$, so it can be used as a non-negative uncertainty measure in subsequent thresholds. The total uncertainty for a user $u$ is aggregated as: $\varsigma_u = \frac{1}{|\mathcal{I}_u|} \sum_{i \in \mathcal{I}_u} (\varsigma_v(\mathbf{h}_i) + \varsigma_t(\mathbf{h}_i))$.

\subsection{Uncertainty-guided Hierarchical Intent Generation (UHIG)}
This module builds a personalized intent hierarchy per user. Unlike conventional methods that fix the number of intents, we use Residual Quantization (RQ) to generate hierarchical intent embeddings whose depth is dynamically controlled by the user uncertainty score $\varsigma_u$ from CUSM.

We define $L$ intent codebooks $\{\mathcal{C}_l \in \mathbb{R}^{K_l \times d}\}_{l=1}^L$, one per level. Initializing $\mathbf{r}_u^{(0)} = \mathbf{h}_u$, the generation proceeds iteratively for each level $l \in \{1,\dots,L\}$:
\begin{align}
\mathbf{c}_u^{(l)} &= \underset{\mathbf{c} \in \mathcal{C}_l}{\text{argmin}} \|\mathbf{r}_u^{(l-1)} - \mathbf{c}\|_2^2 \label{eq:quantization} \\
\mathbf{r}_u^{(l)} &= \mathbf{r}_u^{(l-1)} - \mathbf{c}_u^{(l)} \label{eq:residual}
\end{align}
This decomposes the user embedding into a sequence of increasingly fine-grained intent codes. The generation for user $u$ stops at depth $D_u$ when the residual norm falls below an uncertainty-dependent threshold:
\begin{equation}
\|\mathbf{r}_u^{(D_u)}\|_2 < \tau_{D_u} \quad \text{where} \quad \tau_l = \frac{\tau_0}{l} \cdot \varsigma_u
\end{equation}
Here $\tau_0$ is a tunable hyperparameter. High-uncertainty users get larger thresholds and thus shallower (coarser) hierarchies, while low-uncertainty users develop deeper (finer) ones. This stopping rule realizes a per-user adaptive information bottleneck, suppressing noisy fine-grained factors when $\varsigma_u$ is large.

The codebooks are optimized via the standard VQ-VAE objective together with a diversity loss:
\begin{equation}
\mathcal{L}_{Quant} = \sum_{u \in \mathcal{U}} \sum_{l=1}^{D_u} \left( \|\text{sg}(\mathbf{r}_u^{(l-1)}) - \mathbf{c}_u^{(l)}\|_2^2 + \beta_q \|\mathbf{r}_u^{(l-1)} - \text{sg}(\mathbf{c}_u^{(l)})\|_2^2 \right)
\end{equation}
where $\text{sg}(\cdot)$ is the stop-gradient operator and $\beta_q$ is the commitment cost. A diversity loss further encourages separation between intent codes within the same level:
\begin{equation}
\mathcal{L}_{Div} = \sum_{l=1}^L \sum_{\mathbf{c}_i, \mathbf{c}_j \in \mathcal{C}_l, i \neq j} \max(0, \delta - \|\mathbf{c}_i - \mathbf{c}_j\|_2)
\end{equation}

\subsection{Uncertainty-aware Hierarchical Intent Aggregation}
We aggregate the user-specific intent hierarchy $\{\mathbf{c}_u^{(l)}\}_{l=1}^{D_u}$ via an uncertainty-gated attention mechanism that prioritizes coarser intents for uncertain users:
\begin{equation}
\alpha_l = \frac{\exp \left( \frac{\mathbf{h}_u^\top \mathbf{c}_u^{(l)}}{\sqrt{d}} - \gamma \cdot l \cdot \varsigma_u \right)}{\sum_{k=1}^{D_u} \exp \left( \frac{\mathbf{h}_u^\top \mathbf{c}_u^{(k)}}{\sqrt{d}} - \gamma \cdot k \cdot \varsigma_u \right)}
\end{equation}
where $\gamma$ is a scaling factor. The term $\exp(-\gamma \cdot l \cdot \varsigma_u)$ penalizes finer-grained intents (larger $l$) for users with high uncertainty. The aggregated intent representation is $\mathbf{i}_u = \sum_{l=1}^{D_u} \alpha_l \mathbf{c}_u^{(l)}$, and the final user representation fuses collaborative and intent embeddings: $\mathbf{u}_u^* = \mathbf{h}_u + \mathbf{i}_u$.

\subsection{Prediction and Optimization}
The prediction score is $\hat{y}_{ui} = (\mathbf{u}_u^*)^\top \mathbf{h}_i$. We adopt the Bayesian Personalized Ranking (BPR) loss as the recommendation objective:
\begin{equation}
\mathcal{L}_{Rec} = \sum_{(u,i,j) \in \mathcal{D}} -\log \mathrm{sigmoid}(\hat{y}_{ui} - \hat{y}_{uj})
\end{equation}
where $\mathcal{D} = \{(u,i,j) | u \in \mathcal{U}, i \in \mathcal{I}_u, j \in \mathcal{I} \setminus \mathcal{I}_u\}$ is the set of training triplets. We group the auxiliary losses by module:
\begin{align}
\mathcal{L}_{Uncertainty} &= \mathcal{L}_{FM} + \lambda_{cross}\mathcal{L}_{cross} \\
\mathcal{L}_{Intent} &= \mathcal{L}_{Quant} + \lambda_{div}\mathcal{L}_{Div}
\end{align}
The final UHIFlow objective is:
\begin{equation}
\mathcal{L} = \mathcal{L}_{Rec} + \mu_1 \mathcal{L}_{Uncertainty} + \mu_2 \mathcal{L}_{Intent}
\end{equation}
where $\mu_1, \mu_2$ are hyperparameters balancing the two auxiliary losses.

\section{Experiments}

\subsection{Experimental Settings}

\subsubsection{Datasets}
Following \cite{guo2024lgmrec,jiang2024diffmm}, we evaluate on three widely-used multimodal benchmarks under 5-core filtering, all with pre-extracted visual (V) and textual (T) features. The Amazon datasets are from \cite{mcauley2015image}, and TikTok is from \cite{wei2019mmgcn}. Table~\ref{tab:dataset_statistics} summarizes their statistics.

\begin{table}[h]
\centering
\caption{Statistics of the experimental datasets.}
\label{tab:dataset_statistics}
\begin{adjustbox}{max width=0.82\textwidth}
\begin{tabular}{lccc}
\toprule
\rowcolor{tableheadercolor}
\textbf{Statistics} & \textbf{Amazon-Baby} & \textbf{Amazon-Sports} & \textbf{TikTok} \\
\midrule
\#Users & 19{,}445 & 35{,}598 & 9{,}319 \\
\rowcolor{lightyellow}
\#Items & 7{,}050 & 18{,}357 & 6{,}710 \\
\#Interactions & 139{,}110 & 256{,}308 & 59{,}541 \\
\rowcolor{lightyellow}
Sparsity & 99.899\% & 99.961\% & 99.904\% \\
\bottomrule
\end{tabular}
\end{adjustbox}
\end{table}

\subsubsection{Baseline Methods}
We compare \textsc{UHIFlow} against representative baselines grouped by modeling principle: \textbf{(1) Traditional CF}: NGCF~\cite{wang2019neural}, LightGCN~\cite{he2020lightgcn}; \textbf{(2) Intent-based}: LIP~\cite{li2025lip}, HIM~\cite{Zhu2022HIM}, DCCF~\cite{ren2023disentangled}, BIGCF~\cite{zhang2024exploring}, MMIL~\cite{yang2024multimodal}; \textbf{(3) Uncertainty-aware}: WaPOIR~\cite{Zhou2023WaPOIR}, UCC~\cite{Liu2023UCC}, UA-GNN~\cite{Cao2024UAGNN}; \textbf{(4) Generative}: FlowCF~\cite{liu2025flow}, DiffMM~\cite{jiang2024diffmm}, DiffCL~\cite{song2025diffcl}; and \textbf{(5) Multimodal}: LGMRec~\cite{guo2024lgmrec}, MENTOR~\cite{xu2025mentor}, MIG-GT~\cite{hu2025modality}, D-DPDG~\cite{wu2026ddpdg}, MoToRec~\cite{liu2026motorec}.


\subsubsection{Evaluation Metrics}
Similar to other works in multimodal recommendation \cite{guo2024lgmrec,xu2025mentor}, we adopt two widely-used metrics to evaluate top-K recommendation performance: Recall@K and Normalized Discounted Cumulative Gain (NDCG@K). We report results for K=\{10, 20\}. For each user, we rank all items they have not interacted with in the training set and evaluate the ranking performance on the test set.

\subsubsection{Implementation Details}
We implement UHIFlow in PyTorch and run all experiments on NVIDIA A100 GPUs. For fairness, all models use embedding size $d=64$; UHIFlow is trained with Adam using learning rate $10^{-3}$, weight decay $10^{-4}$, and batch size 2048. We use $K=3$ graph propagation layers, a two-layer MLP vector field with hidden size 64 for CUSM, an adaptive dopri5 ODE solver with relative tolerance $1e-5$, a single linear cross-modal projection head, and UHIG depth $L=4$ with codebook sizes $\{128,256,512,1024\}$ and commitment cost 0.25. We tune $\mu_1,\mu_2 \in \{0.01,\dots,1.0\}$, $\lambda_{cross} \in \{0.1,\dots,2.0\}$, and $\lambda_{div} \in \{0.001,\dots,0.1\}$ by grid search; all baselines are run with official code and fine-tuned parameters, and results are averaged over 5 random seeds.

\subsection{Overall Performance Comparison}

The main results are presented in Table \ref{tab:main_results}. We summarize the key observations below.

\begin{table}[h]
\centering
\caption{Overall performance comparison on the three datasets. \textbf{Bold} indicates the best and \underline{underline} indicates the second-best result. `Improv.' denotes the relative improvement of UHIFlow over the best baseline. All improvements are significant ($p < 0.05$).}
\renewcommand{\arraystretch}{1.08}
\label{tab:main_results}
\tabvspace
\large
\setlength{\tabcolsep}{2pt}
\begin{adjustbox}{max width=\textwidth}
\begin{tabular}{@{}l l cccc cccc cccc@{}}
\toprule
\rowcolor{headergray}
\multirow{2}{*}{\textbf{Category}} & \multirow{2}{*}{\textbf{Method}} & \multicolumn{4}{c}{\textbf{Amazon-Baby}} & \multicolumn{4}{c}{\textbf{Amazon-Sports}} & \multicolumn{4}{c}{\textbf{TikTok}} \\ \cmidrule(lr){3-6} \cmidrule(lr){7-10} \cmidrule(lr){11-14}
\rowcolor{headergray}
& & R@10 & R@20 & N@10 & N@20 & R@10 & R@20 & N@10 & N@20 & R@10 & R@20 & N@10 & N@20 \\ \midrule \midrule
\multirow{2}{*}{Traditional CF}
& NGCF & 0.0388 & 0.0591 & 0.0208 & 0.0261 & 0.0468 & 0.0695 & 0.0250 & 0.0318 & 0.0366 & 0.0604 & 0.0159 & 0.0238 \\
& LightGCN & 0.0479 & 0.0754 & 0.0257 & 0.0328 & 0.0569 & 0.0864 & 0.0311 & 0.0387 & 0.0396 & 0.0653 & 0.0189 & 0.0282 \\
\midrule

\multirow{3}{*}{Uncertainty}
& WaPOIR & 0.0496 & 0.0752 & 0.0260 & 0.0325 & 0.0588 & 0.0868 & 0.0309 & 0.0375 & 0.0412 & 0.0672 & 0.0177 & 0.0270 \\
& UCC & 0.0512 & 0.0780 & 0.0271 & 0.0340 & 0.0605 & 0.0897 & 0.0321 & 0.0391 & 0.0420 & 0.0689 & 0.0186 & 0.0280 \\
& UA-GNN & 0.0531 & 0.0809 & 0.0281 & 0.0353 & 0.0622 & 0.0923 & 0.0330 & 0.0402 & 0.0432 & 0.0708 & 0.0190 & 0.0287 \\
\midrule

\multirow{5}{*}{Intent-based}
& LIP & 0.0617 & 0.0952 & 0.0336 & 0.0427 & 0.0702 & 0.1055 & 0.0383 & 0.0472 & 0.0576 & 0.0962 & 0.0275 & 0.0402 \\
& HIM & 0.0556 & 0.0858 & 0.0303 & 0.0384 & 0.0656 & 0.0986 & 0.0358 & 0.0442 & 0.0454 & 0.0758 & 0.0217 & 0.0317 \\
& DCCF & 0.0547 & 0.0848 & 0.0301 & 0.0383 & 0.0648 & 0.0978 & 0.0356 & 0.0441 & 0.0443 & 0.0744 & 0.0216 & 0.0314 \\
& BIGCF & 0.0574 & 0.0882 & 0.0310 & 0.0392 & 0.0672 & 0.1005 & 0.0363 & 0.0447 & 0.0477 & 0.0792 & 0.0222 & 0.0328 \\
& MMIL & 0.0601 & 0.0916 & 0.0319 & 0.0399 & 0.0678 & 0.1006 & 0.0359 & 0.0438 & 0.0531 & 0.0871 & 0.0235 & 0.0354 \\
\midrule

\multirow{3}{*}{Generative}
& FlowCF & 0.0561 & 0.0870 & 0.0309 & 0.0393 & 0.0662 & 0.1000 & 0.0364 & 0.0452 & 0.0468 & 0.0787 & 0.0229 & 0.0332 \\
& DiffMM & 0.0623 & 0.0975 & 0.0328 & 0.0411 & 0.0683 & 0.1019 & 0.0374 & 0.0455 & 0.0684 & \second{0.1129} & \second{0.0306} & \second{0.0456} \\
& DiffCL & 0.0641 & 0.0987 & 0.0343 & 0.0433 & 0.0754 & 0.1095 & \second{0.0421} & 0.0509 & 0.0677 & 0.1112 & 0.0299 & 0.0451 \\
\midrule

\multirow{5}{*}{Multimodal}
& LGMRec & 0.0644 & 0.1002 & 0.0349 & 0.0440 & 0.0720 & 0.1068 & 0.0390 & 0.0480 & 0.0654 & 0.1072 & 0.0288 & 0.0435 \\
& D-DPDG & 0.0655 & 0.0993 & 0.0344 & 0.0429 & 0.0742 & 0.1096 & 0.0389 & 0.0473 & 0.0674 & 0.1098 & 0.0289 & 0.0441 \\
& MIG-GT & 0.0665 & 0.1021 & 0.0361 & \second{0.0452} & 0.0753 & 0.1130 & 0.0414 & 0.0510 & 0.0678 & 0.1105 & 0.0292 & 0.0444 \\
& MENTOR & \second{0.0678} & \second{0.1048} & \second{0.0362} & 0.0450 & \second{0.0763} & \second{0.1139} & 0.0409 & \second{0.0511} & \second{0.0691} & 0.1120 & 0.0290 & 0.0446 \\
& MoToRec & 0.0676 & 0.1027 & 0.0358 & 0.0451 & 0.0747 & 0.1109 & 0.0413 & 0.0504 & 0.0648 & 0.1070 & 0.0293 & 0.0436 \\
\midrule
\textbf{Ours} & \textbf{UHIFlow} & \best{0.0698} & \best{0.1079} & \best{0.0373} & \best{0.0465} & \best{0.0786} & \best{0.1171} & \best{0.0433} & \best{0.0526} & \best{0.0712} & \best{0.1162} & \best{0.0315} & \best{0.0469} \\
\midrule
\rowcolor{lightyellow}
\multicolumn{2}{@{}l}{\textbf{Improv.}} & 2.95\% & 2.96\% & 3.04\% & 2.88\% & 3.01\% & 2.81\% & 2.85\% & 2.94\% & 3.04\% & 2.92\% & 2.94\% & 2.85\% \\
\bottomrule
\end{tabular}
\end{adjustbox}
\end{table}

UHIFlow achieves the best performance across all datasets and metrics, improving over the strongest baseline (MENTOR, AAAI'25) by 2.81\%--3.04\%. The gains are consistent on both sparse Amazon catalogs and the noisier TikTok micro-video domain, and UHIFlow also surpasses recent competitors such as MoToRec and D-DPDG on every metric. This supports our core design: modeling multimodal uncertainty and adaptive hierarchical intent jointly provides a stable signal, allowing uncertain cases to back off to coarse intents while confident cases refine toward item-level granularity.

Category-level results further show that advanced multimodal, intent-based, and generative methods outperform traditional CF. MENTOR and MIG-GT are strong multimodal competitors, and LIP is the strongest intent-based baseline, but they still rely on flat or fixed intent structures. FlowCF, DiffMM, and DiffCL remain competitive, yet mainly use flow or diffusion for denoising rather than for \emph{quantifying} residual uncertainty. UHIFlow closes these gaps by coupling uncertainty estimation, cross-modal synergy, and user-specific hierarchy construction.

\subsection{Ablation Study}
To dissect our model and understand the contribution of its key components, we conduct an ablation study with several variants of UHIFlow. The results are presented in Table \ref{tab:ablation_study}.
\begin{itemize}
    \item \textbf{w/o CIP}: We remove the Collaborative Interest Propagation layers and use only the initial multimodal features.
    \item \textbf{w/o UHIG}: We remove the entire Uncertainty-guided Hierarchical Intent Generation module. The final user representation is simply the GNN embedding $\mathbf{h}_u$.
    \item \textbf{w/o CUSM}: We remove the Cross-modal Uncertainty Synergistic Modeling module. Consequently, user uncertainty is unavailable, and we use a fixed-depth hierarchy ($D_u=2$) for all users.
    \item \textbf{w/o cross}: We disable the cross-modal synergistic loss ($\mathcal{L}_{cross}$) in CUSM, so uncertainties are estimated independently.
    \item \textbf{w/o hierarchy}: We replace the hierarchical intent generation with a flat, single-level intent codebook.
\end{itemize}

\begin{table}[h]
\centering
\caption{Ablation study of UHIFlow's core components on three datasets, reporting Recall@20. Performance drops from the full model are highlighted in \textcolor{red}{red}.}
\label{tab:ablation_study}
\tabvspace
\scriptsize
\renewcommand{\arraystretch}{1.15}
\setlength{\tabcolsep}{2pt}
\setlength{\aboverulesep}{0.35ex}
\setlength{\belowrulesep}{0.45ex}

\begin{adjustbox}{max width=0.9\textwidth}
    \begin{tabular}{l c c c}
    \toprule
    \rowcolor{lightblue}
    \textbf{Variant} & \textbf{Amazon-Baby} & \textbf{Amazon-Sports} & \textbf{TikTok} \\
    \midrule
    \rowcolor{lightyellow}
    \textbf{UHIFlow (Full Model)} & \textbf{0.1079} & \textbf{0.1171} & \textbf{0.1162} \\
    \midrule
    w/o CIP & 0.0981 (\textcolor{red}{-0.0098}) & 0.1049 (\textcolor{red}{-0.0122}) & 0.1056 (\textcolor{red}{-0.0106}) \\

    w/o UHIG & 0.1008 (\textcolor{red}{-0.0071}) & 0.1079 (\textcolor{red}{-0.0092}) & 0.1085 (\textcolor{red}{-0.0077}) \\

    w/o CUSM & 0.1015 (\textcolor{red}{-0.0064}) & 0.1092 (\textcolor{red}{-0.0079}) & 0.1093 (\textcolor{red}{-0.0069}) \\

    w/o cross & 0.1060 (\textcolor{red}{-0.0019}) & 0.1146 (\textcolor{red}{-0.0025}) & 0.1119 (\textcolor{red}{-0.0043}) \\

    w/o hierarchy & 0.1040 (\textcolor{red}{-0.0039}) & 0.1127 (\textcolor{red}{-0.0044}) & 0.1106 (\textcolor{red}{-0.0056}) \\
    
    \bottomrule
    \end{tabular}
\end{adjustbox}
\end{table}

Removing any component degrades performance. The largest drops occur for \textbf{w/o CIP} and \textbf{w/o UHIG}, highlighting the necessity of collaborative propagation and intent modeling, while the decline for \textbf{w/o CUSM} confirms that uncertainty-guided adaptation beats a fixed strategy. Each component, from the base GNN to synergistic uncertainty modeling and hierarchical generation, thus plays an integral role in UHIFlow. The two core modules are moreover complementary: CUSM supplies the per-user uncertainty that UHIG consumes to set its depth, so ablating CUSM not only removes the uncertainty signal but also collapses the hierarchy to a fixed granularity---which is why its drop is larger than that of the cross-modal term alone.

\subsection{Impact of Uncertainty Modeling Strategy}

\begin{table}[h]
    \centering
    \caption{Performance comparison.}
    \label{tab:uncertainty_variants}
    \tabvspace
    \scriptsize
    \renewcommand{\arraystretch}{1.15}
    \setlength{\tabcolsep}{2pt}
    \setlength{\aboverulesep}{0.35ex}
    \setlength{\belowrulesep}{0.45ex}
    
    \begin{adjustbox}{max width=0.82\textwidth}
        \begin{tabular}{l c c c c}
        \toprule
        \rowcolor{tableheadercolor} 
         & \multicolumn{2}{c}{\textbf{Amazon-Baby}} & \multicolumn{2}{c}{\textbf{Amazon-Sports}} \\
        \cmidrule(lr){2-3} \cmidrule(lr){4-5} 
        \rowcolor{tableheadercolor}
        \multirow{-2}{*}{\textbf{Strategy}} & \textbf{R@20} & \textbf{N@20} & \textbf{R@20} & \textbf{N@20} \\
        
        \midrule
        Gaussian Embedding & 0.1032 & 0.0443 & 0.1129 & 0.0500 \\
    
        Wasserstein Distance & 0.1048 & 0.0452 & 0.1143 & 0.0509 \\
    
        Energy-Based Model & 0.1064 & 0.0458 & 0.1156 & 0.0516 \\
        \rowcolor{ourscolor}
        \textbf{Flow Matching (Ours)} & \textbf{0.1079} & \textbf{0.0465} & \textbf{0.1171} & \textbf{0.0526} \\
        
        \bottomrule
        \end{tabular}
    \end{adjustbox}
    \end{table}

To demonstrate the superiority of Flow Matching for uncertainty quantification, we compare UHIFlow with variants employing alternative uncertainty modeling strategies:
\begin{itemize}
    \item \textbf{Gaussian Embedding (GD)}: Models feature uncertainty by mapping items to Gaussian distributions $\mathcal{N}(\mu, \Sigma)$ and using KL divergence for alignment.
    \item \textbf{Wasserstein Distance (WD)}~\cite{Zhou2023WaPOIR}: Uses Wasserstein distance to measure the discrepancy between distributions, as used in WaPOIR.
    \item \textbf{Energy-Based Model (EBM)}: Quantifies uncertainty via energy scores derived from an implicit energy function.
\end{itemize}

Table \ref{tab:uncertainty_variants} shows that \textbf{Flow Matching} consistently outperforms other strategies. Gaussian and Wasserstein methods improve over deterministic baselines but rely on restrictive assumptions (e.g., Gaussianity), while EBMs are more flexible yet unstable to train. Flow Matching strikes the best balance, giving stable, expressive, and accurate uncertainty quantification for hierarchical intent generation.

\subsection{Parameter Sensitivity Analysis}

\begin{figure}[h]
    \centering
    \includegraphics[width=0.8\linewidth]{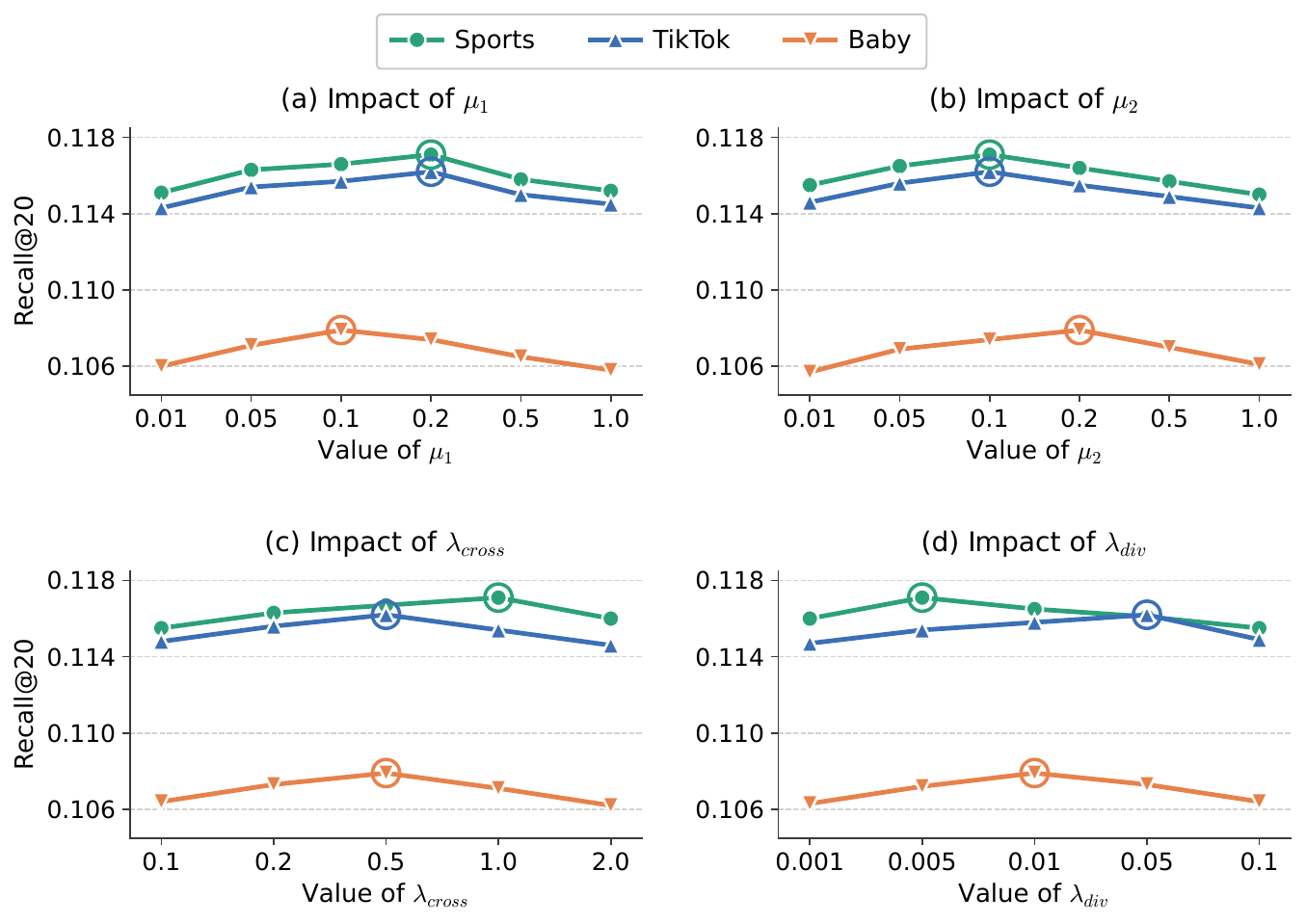}
    \caption{Hyperparameter sensitivity analysis of UHIFlow on the Baby, Sports, and TikTok datasets. We report Recall@20 as we vary the values of four key hyperparameters: (a) uncertainty loss weight $\mu_1$, (b) intent loss weight $\mu_2$, (c) cross-modal guidance weight $\lambda_{cross}$, and (d) intent diversity weight $\lambda_{div}$.}
    \label{fig:hyper_sensitivity}
\figvspace
\end{figure}


We study four key hyperparameters: the loss weights $\mu_1$ (uncertainty), $\mu_2$ (intent), $\lambda_{cross}$ (cross-modal guidance), and $\lambda_{div}$ (intent diversity), varying one at a time while keeping the others optimal. As Figure \ref{fig:hyper_sensitivity} reports on Recall@20, performance first rises then falls as each weight increases, reflecting a trade-off between the auxiliary tasks and the main objective. UHIFlow stays robust across a reasonable range (e.g., $\mu_1, \mu_2 \in [0.1, 0.5]$), confirming its stability and practicality.

\subsection{Efficiency Analysis}

\begin{figure}[h]
    \centering
    \includegraphics[width=0.8\columnwidth]{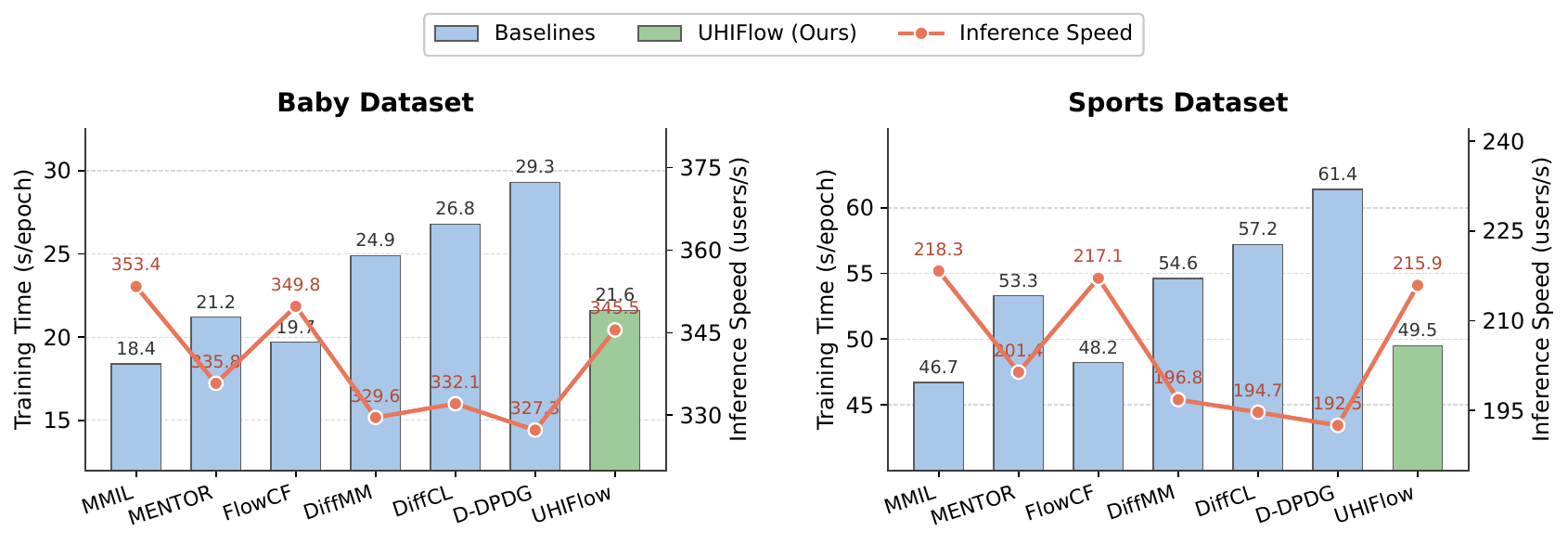}
    \caption{Efficiency comparison on the Baby and Sports datasets. Bars represent training time per epoch (left y-axis, lower is better), and the line represents inference speed (right y-axis, higher is better).}
    \label{fig:efficiency}
\figvspace
\end{figure}

We assess UHIFlow's computational efficiency against representative non-generative baselines (MMIL, MENTOR) and same-category generative baselines---the flow-based FlowCF and the diffusion-based DiffMM, DiffCL, and D-DPDG---on an NVIDIA A100 GPU. As shown in Figure \ref{fig:efficiency}, the diffusion-based methods (DiffMM, DiffCL, D-DPDG) incur the highest training and inference cost because of their iterative multi-step denoising. In contrast, UHIFlow's flow matching introduces only a moderate training overhead---on par with the non-generative models and consistently lower than every diffusion-based competitor---while its inference speed stays close to the fastest baselines, as the expensive flow computation is performed offline during training. This favorable balance between accuracy and computational cost confirms UHIFlow's suitability for practical deployment.



\section{Conclusion}
\label{sec:conclusion}

We proposed UHIFlow to address the inherent uncertainty and static intent structures of multimodal recommendation. Its Cross-modal Uncertainty Synergistic Modeling (CUSM) module uses conditional flow matching to quantify and synergize multimodal uncertainties, which then adaptively control the depth of an Uncertainty-guided Hierarchical Intent Generation (UHIG) module that builds personalized, fine-to-coarse intent structures per user. Extensive experiments demonstrate UHIFlow's superiority over state-of-the-art baselines. Future work will explore sample-efficient uncertainty estimation from limited-feedback signals~\cite{miao2026universal}.

\begin{credits}
\subsubsection{\ackname}
This work was supported by the Open Project Program of Marine Ecological Restoration and Smart Ocean Engineering Research Center of Hebei Province under Grant No.~HBMESO2507; the Shijiazhuang--Northeastern University Science and Technology Cooperation Special Project under Grant No.~NEUS2025-01-003; the Scientific Research Project of Hebei Education Department under Grant No.~QN2024167; and the National College Student Innovation and Entrepreneurship Training Program under Project No.~202619145026.
\end{credits}

\bibliographystyle{splncs04}
\bibliography{references}

\end{document}